\documentclass[12pt]{article}

\usepackage{cjp}
\usepackage{epsfig}

\newcommand{\Sign}{\mbox{Sign}}

\newcommand{\B}{{\mathrm{b}}}
\newcommand{\e}{{\mathrm{e}}}
\newcommand{\f}{{\mathrm{f}}}

\newcommand{\bSign}{{\overline{\Sign[{\cal C}]}}}

\begin{document}
\title{A Chiral Phase Transition using a \\
Fermion Cluster Algorithm}
\author{Shailesh Chandrasekharan}
\address{Department of Physics, Duke University \\
         Durham, NC 27708-0305,  USA. }
\date{\today}
\maketitle
\begin{abstract}
The recent solution to the fermion sign problem allows, for the first time, 
the use of cluster algorithm techniques to compute certain fermionic 
path integrals. To illustrate the underlying ideas behind the progress,
a cluster algorithm is constructed to study the chiral phase transition in 
a strongly interacting staggered fermion model with an arbitrary 
mass term in $3+1$ dimensions. Unlike conventional 
methods there is no difficulty in the cluster method to approach the chiral 
(massless) limit. Results using the new algorithm confirm that the chiral 
transition falls under the expected universality class.
\end{abstract}
\section{Introduction}

The past few years has been a time of remarkable progress in our
understanding of lattice fermions and their ability to reproduce
the chiral properties of the continuum theory \cite{Neu99}. It is
now possible to formulate vector-like gauge theories on the 
lattice with all its chiral symmetry intact \cite{Cha98}. Our
understanding of non-perturbative formulations of chiral gauge 
theories is also advancing at a remarkable rate as is reflected by 
many recent contributions \cite{Lus99}. On the other hand, our 
ability to perform reliable (numerical) calculations in many theories 
involving strongly interacting fermions and especially in the chiral 
limit has remained primitive. In particular algorithms based
on hybrid molecular dynamics, that are popular for gauge theories like
QCD, suffer either from sign problems or critical slowing down in the chiral 
limit. Unless better numerical methods are found, it is likely that we 
will be forced to perform most of our calculations far from the chiral 
limit and hence not gain from the progress in our understanding of chiral 
symmetry.

Numerical methods for fermionic systems are especially difficult 
due to the Pauli-exclusion principle. The Boltzmann weight of the 
statistical mechanics problem describing such systems can often 
become negative thereby invalidating most Monte-Carlo methods. 
This is usually referred to as the sign problem. In other cases 
one is often forced to work with bosonic variables with non-local 
Boltzmann weights which arise through the fermion determinant. 
Such expressions are not easy to handle and there is no known way 
to overcome problems like critical slowing.

Recently a new numerical method to perform fermionic path integrals 
has emerged and is based on a rather novel form of bosonization. Instead 
of the Grassmann variables, the method starts by representing fermions
through their occupation numbers. Such representations, which have been 
studied in the past \cite{Dun88,Wie93}, were not considered useful due to
sign problems. Over the past year it has become clear 
that at least in a class of models it is possible to design cluster 
algorithms which automatically solve the fermion sign problem and
provide an alternative way to compute fermionic path integrals. In a
sense the algorithm suggests a new way to bosonize the model in terms 
of dynamics of clusters. The new method works in both relativistic
\cite{Sch99.1,Sch99.2} and non-relativistic\cite{Sch99.3} models.
Since cluster algorithms are known to beat critical slowing down
in a variety of bosonic models\cite{Eve93,Eve97}, the new algorithms 
have the ability to work efficiently even in the presence of long 
correlation lengths. In particular, they encounter no difficulties in 
the limit where the fermions become massless.

The present article illustrates the main features of the new approach through 
an example of a four Fermi model involving staggered fermions. This model 
undergoes a $Z_2$ chiral transition from a high temperature chirally 
symmetric phase to the broken phase at low temperatures and was studied 
originally in the chiral limit in \cite{Sch99.2}. Here the earlier work is 
extended to include a fermion mass. This extension turns out to be quite 
useful because on a large but finite lattice all the interesting properties 
of the chiral condensate, as an order parameter of the transition, emerge 
only when a tiny mass can be added to the system. In the next section the 
partition function of the model is written in the occupation number basis 
and the origin of the fermion sign problem is reviewed. In section 3 the 
partition function is rewritten in terms of connected fermion world-line
configurations that arise due to the introduction of local ``bond'' 
variables. In the new variables the solution to the sign problem emerges 
naturally and leads to a very elegant cluster algorithm. In section 4, the 
algorithm and some numerical results are discussed. Section 5 contains 
conclusions along with a description of on-going work and points to new 
directions for the future.

\section{The Model}

The model considered in this article involves interacting
staggered fermions hopping on a 3-d cubic spatial lattice 
with $V = L^3$ sites $x$ ($L$ even) and with anti-periodic 
spatial boundary conditions. The dynamics of the fermions 
is described through the Hamilton operator
\begin{equation}
H = \sum_{x,i} h_{x,i} + m \sum_x p_x
\end{equation}
where the term 
\begin{equation}
h_{x,i} = \frac{1}{2} \eta_{x,i}(\Psi_x^+ \Psi_{x+\hat i} + 
\Psi_{x+\hat i}^+ \Psi_x) + (\Psi_x^+ \Psi_x - \frac{1}{2})
(\Psi_{x+\hat i}^+ \Psi_{x+\hat i} - \frac{1}{2}),
\end{equation}
couples the fermion operators at the lattice sites $x$ and $x+\hat i$, 
where $\hat i$ is a unit-vector in the $i$-direction and the mass 
term 
\begin{equation}
p_x =  (-1)^{x_1+x_2+x_3} (\Psi_x^+ \Psi_x-1/2),
\end{equation}
is a single site operator. The fermion creation and annihilation operators 
$\Psi_x^+$ and $\Psi_x$ used in the above equations satisfy 
the standard anti-commutation relations
\begin{equation}
\{\Psi_x^+,\Psi_y^+\} = \{\Psi_x,\Psi_y\} = 0, \, 
\{\Psi_x^+,\Psi_y\} = \delta_{xy}.
\end{equation}
Further $\eta_{x,1} = 1$, $\eta_{x,2} = (-1)^{x_1}$ and 
$\eta_{x,3} = (-1)^{x_1 + x_2}$ are the standard staggered fermion 
sign factors. In the chiral limit ($m=0$), the above Hamiltonian
was discussed in \cite{Sch99.2}. 

In addition to the usual $U(1)$ fermion number symmetry, there are a 
number of discrete symmetries of the staggered fermion Hamiltonian
as discussed in \cite{Sus77}. For example, at $m=0$ the Hamiltonian is 
invariant under shifts of $\Psi_x^+$ and $\Psi_x$ by one lattice spacing 
in the $x_3$ direction. The mass term breaks this $\mathbf{Z}_2$ symmetry 
which is known to be a subgroup of the well known chiral symmetry of 
relativistic massless fermions. Here this symmetry is broken spontaneously 
at low temperatures while thermal fluctuations restore it at high 
temperatures.  The associated second order phase transition belongs to 
the universality class of the 3d ising model as discussed in \cite{Sch99.2}. 
The extension discussed here, allows one to add a tiny mass
and hence study the transition using the chiral condensate.

In order to construct the path integral representation of the
partition function the Hamilton operator is first decomposed
into seven terms $H = H_1 + H_2 + ... + H_6 + H_7$ with
\begin{equation}
H_i = \!\!\! \sum_{\stackrel{x = (x_1,x_2,x_3)}{x_i even}} \!\!\! h_{x,i}
\, \,\;\mbox{and}\;
H_{i+3} = \!\!\! \sum_{\stackrel{x = (x_1,x_2,x_3)}{x_i odd}} \!\!\! h_{x,i}
\,\,\;\mbox{for $i=1,2,3$};\;\;\; H_7 = m \sum_x p_x.
\end{equation}
This decomposition is exactly the same as the one used in \cite{Sch99.2}
for the $m=0$ case, where $H_7$ was absent. The various steps in the 
construction of the partition function from here on is exactly the same
as in \cite{Sch99.2}. Leaving the details to that paper only the
essential points are sketched here. The Suzuki-Trotter formula leads to 
the expression 
\begin{equation}
Z_f = \mbox{Tr} [\exp(- \beta H)] = \lim_{M \rightarrow \infty} \mbox{Tr} 
[\e^{- \epsilon H_1} \e^{- \epsilon H_7/6} \e^{- \epsilon H_2} 
\e^{- \epsilon H_7/6}... \e^{- \epsilon H_6}\e^{- \epsilon H_7/6}]^M,
\end{equation}
for the fermionic partition function at inverse temperature $\beta$,
where $\epsilon = \beta/M$  is the lattice spacing in the Euclidean time 
direction. Spreading $H_7$ symmetrically with each $H_i,i=1,..6$
is not necessary except that it adds some symmetry to the algorithm. 
Further, in the arguments given 
below, $M$ is assumed to be finite and that the true partition function is 
obtained from an extrapolation of the data obtained from a series of 
simulations at larger and larger $M$. Fortunately it is also possible to 
formulate the cluster algorithm in the time continuum limit as 
demonstrated in \cite{Bea96}.

\begin{figure}[ht]
\hbox{
\hspace{3cm}
\epsfig{file=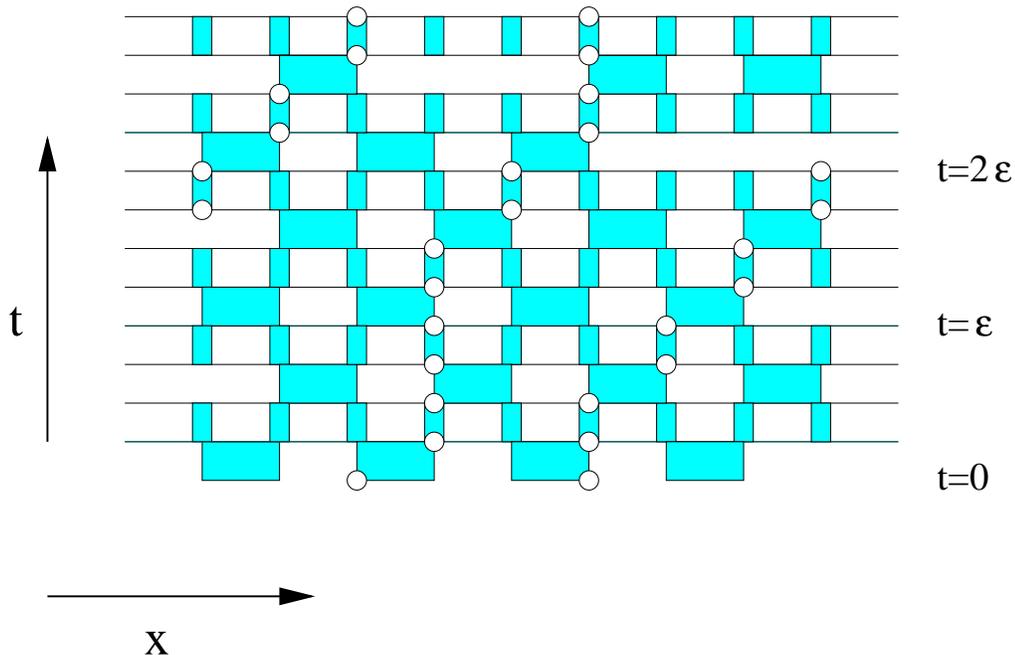, width=5cm,angle=0,
bbllx=50,bblly=50,bburx=225,bbury=300}
}
\vspace{1cm}
\figcaption{\label{fconf}A fermion world-line configuration 
with $\Sign_\f[n] = -1$ on a one dimensional periodic spatial lattice 
of size $L=8$ where the time slice $t=3\varepsilon$ is the same as $t=0$.}
\end{figure}

It is well known that in the occupation number basis a fermionic 
partition function can be written as a sum over fermion world-line 
configurations\cite{Wie93}. Since the fermionic creation operators 
anti-commute,
the information that a subset of the lattice sites are occupied 
defines the state of the system only up to a sign. A complete 
definition of the occupation number basis states requires an ordering 
of the entire lattice which then specifies the order in which the 
fermions on the lattice are created. Taking this ordering into account 
it is possible to write the partition function
as a sum over configurations of fermion occupation numbers $n(x,t) = 0,1$ 
on a $(3+1)$-d space-time lattice of points $(x,t)$ where 
$t=\epsilon n/12, n=0,1,2,...(12M-1)$ labels the time slices. The trace 
imposes the periodicity constraint $n(x,0) = n(x,\beta)$. A typical 
configuration in one spatial dimension is shown in figure 1, where the 
shaded regions depict either a two site interaction due to 
$\exp(-h_{x,i})$ or a single site interaction due to $\exp(-m p_x)$. 

The Boltzmann weight of each configuration $n$ is a product of transfer 
matrix elements associated with the configuration of fermions on each 
shaded region. These elements are shown in figure 2.
\begin{figure}[ht]
\hbox{ \hspace{0.2cm} Non-zero elements due to $\exp(-h_{x,i})$
\hspace{2cm}
Non-zero elements due to $\exp(-p_{x})$ }
\hbox{
\vbox{
\vspace{-4.8cm}
\hbox{
\hspace{1cm}
\epsfig{file=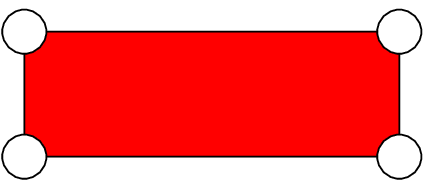,
width=3cm,angle=0,
bbllx=50,bblly=50,bburx=225,bbury=337}
\hspace{1cm}
\epsfig{file=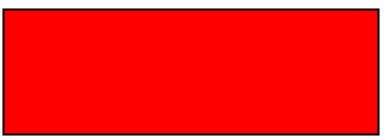,
width=3cm,angle=0,
bbllx=50,bblly=50,bburx=225,bbury=337}
}
\vspace{0.8cm}
\hbox{ \hspace{1cm} (a) \hspace{3.2cm} (b) }
\vspace{-4.5cm}
\hbox{
\hspace{1cm}
\epsfig{file=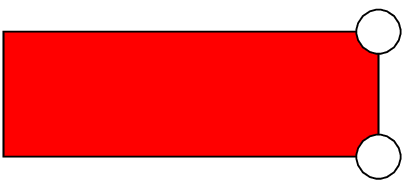,
width=3cm,angle=0,
bbllx=50,bblly=50,bburx=225,bbury=337}
\hspace{0.8cm}
\epsfig{file=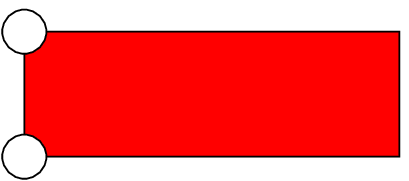,
width=3cm,angle=0,
bbllx=50,bblly=50,bburx=225,bbury=337}
}
\vspace{0.8cm}
\hbox{ \hspace{1cm} (c) \hspace{3.2cm} (d) }
\vspace{-4.5cm}
\hbox{
\hspace{1cm}
\epsfig{file=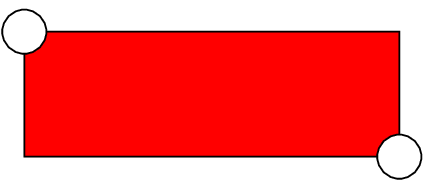,
width=3cm,angle=0,
bbllx=50,bblly=50,bburx=225,bbury=337}
\hspace{0.8cm}
\epsfig{file=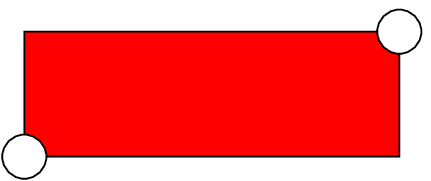,
width=3cm,angle=0,
bbllx=50,bblly=50,bburx=225,bbury=337}
}
\vspace{0.8cm}
\hbox{ \hspace{1cm} (e) \hspace{3.2cm} (f) }
}
\vbox{
\vspace{-4.5cm}
\hbox{
\hspace{1cm}
\epsfig{file=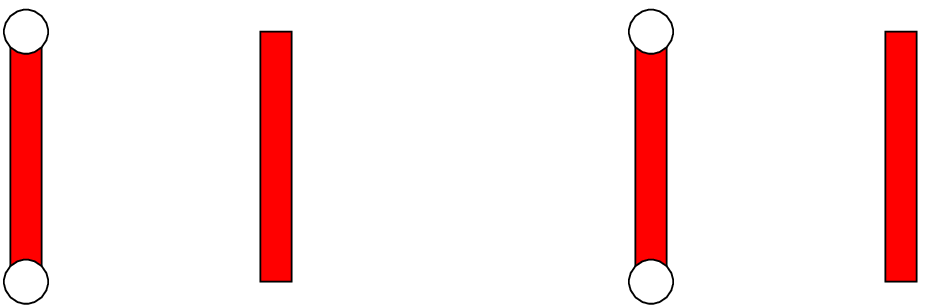,
width=4cm,angle=0,
bbllx=50,bblly=50,bburx=225,bbury=337}
}
\vspace{1.5cm}
\hbox{ { \hspace{-0.3cm} (g) \hspace{0.8cm} (h) \hspace{1.8cm} (i)
\hspace{0.8cm} (j) } }
\vspace{0.5cm}
\hbox{ { \hspace{0.05cm} even site \hspace{2.3cm} odd site } }
\vspace{0.5cm}
}
}
\figcaption{ \label{tmat}The figures (a) through (j) illustrate the
non-zero transfer matrix elements. Weight of (a),(b) is $\exp(-\epsilon/4)$; 
weight of (c),(d) is $\exp(\epsilon/4)\cosh(\epsilon/2)$;
weight of (e),(f) is $\Sigma \exp(\epsilon/4)\sinh(\epsilon/2)$;
weight of (g),(j) is $\exp(-m\epsilon/12)$ and 
weight of (h),(j) is $\exp(m\epsilon/12)$. The factor $\Sigma$
is a product of local sign factors $\eta_{x,i}$ and non-local
sign factors that arise due to anti-commutation relations. Further,
$x_1+x_2+x_3 =$ even(odd) defines an even(odd) site.}
\end{figure}
Representing the magnitude of the Boltzmann weight by $W[n]$ and 
the sign, which takes into account the various anti-commutation relations,
by $\Sign[n]$ the partition function takes the form
\begin{equation}
\label{zfn}
Z_f = \sum_n \Sign[n] W[n].
\end{equation}
Clearly $\Sign[n]$ is the product of the sign factors $\Sigma$ that appear 
in the transfer matrix elements associated to fermion hops represented 
through figures 2(e) and 2(f). When the fermion hops from $x$ 
to $x+\hat{i}$ or vice versa due to the action of  $\exp(-h_{x,i})$ it 
picks up a product of local sign factors due to terms like $\eta_{x,i}$ 
as well as a string of non-local signs that arise due to anti-commutation
relations involved when the fermion has to cross other fermions on the 
ordered lattice while reaching its destination. The evaluation of this 
non-local part of $\Sigma$ is rather tedious. Separating the two parts, 
one can define $\Sign[n]=\Sign_\f[n]\Sign_\B[n]$, where $\Sign_f[n]$ comes 
from the product of non-local parts of $\Sigma$ and is purely fermionic, 
where as $\Sign_\B[n]$ is the one that comes from the local parts and 
may arise even in bosonic models. Fortunately $\Sign_\f[n]$ has a 
topological meaning. The occupied lattice sites define fermion world-lines 
which are closed around the Euclidean time direction. However, during 
their Euclidean time evolution fermions can interchange their positions, 
and the fermion world-lines thus define a permutation of particles. The 
Pauli exclusion principle dictates that the $\Sign_f[n]$ is just the sign 
of that permutation. Further, $\Sign_\B[n]$ receives an extra minus-sign 
for every fermion that hops across a spatial boundary due to anti-periodic 
boundary conditions, 

\begin{figure}[htb]
\hbox{
\hspace{0.7cm}
\epsfig{file=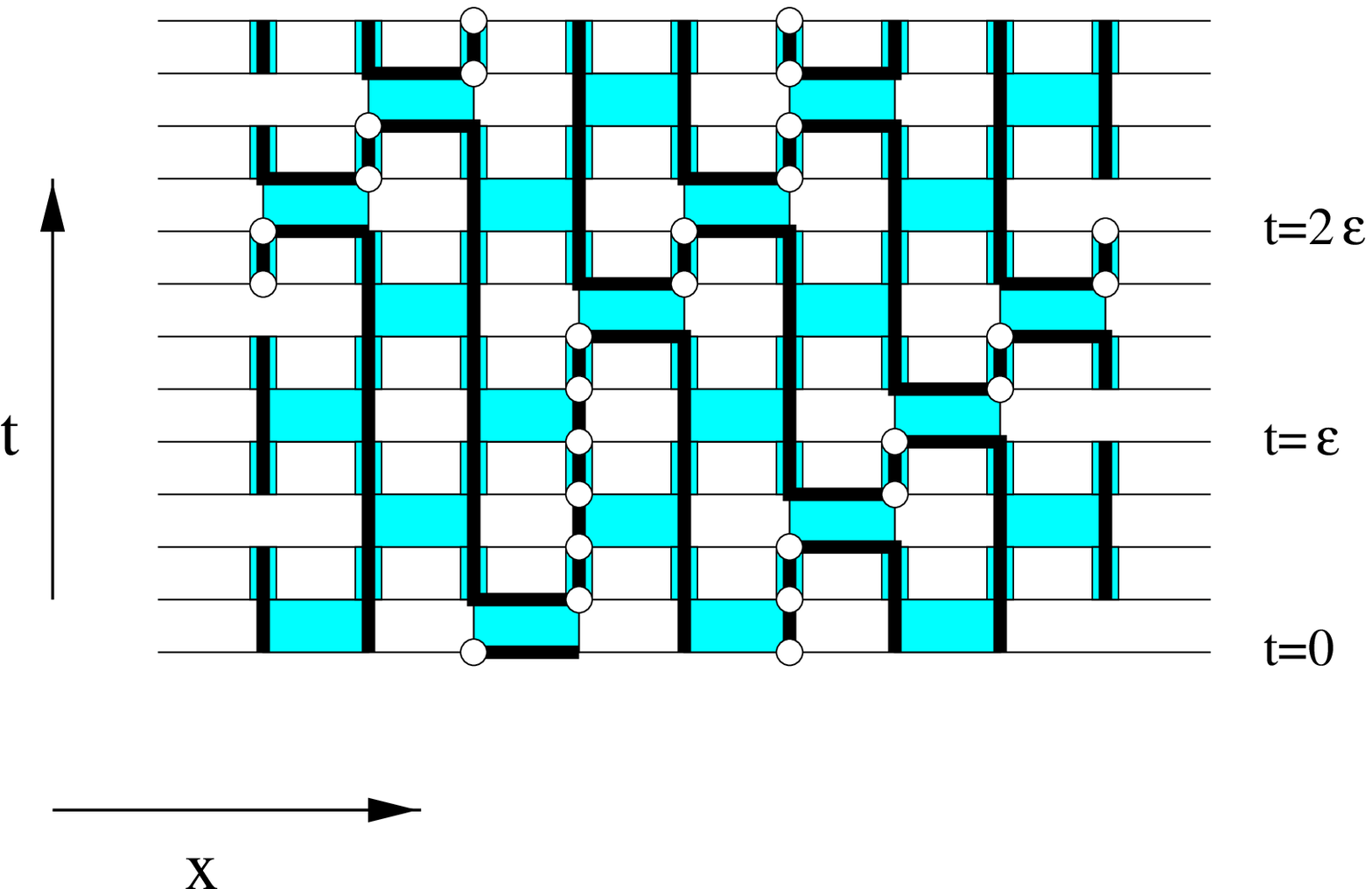,
width=3cm,angle=0,
bbllx=50,bblly=50,bburx=225,bbury=337}
\hspace{4.7cm}
\epsfig{file=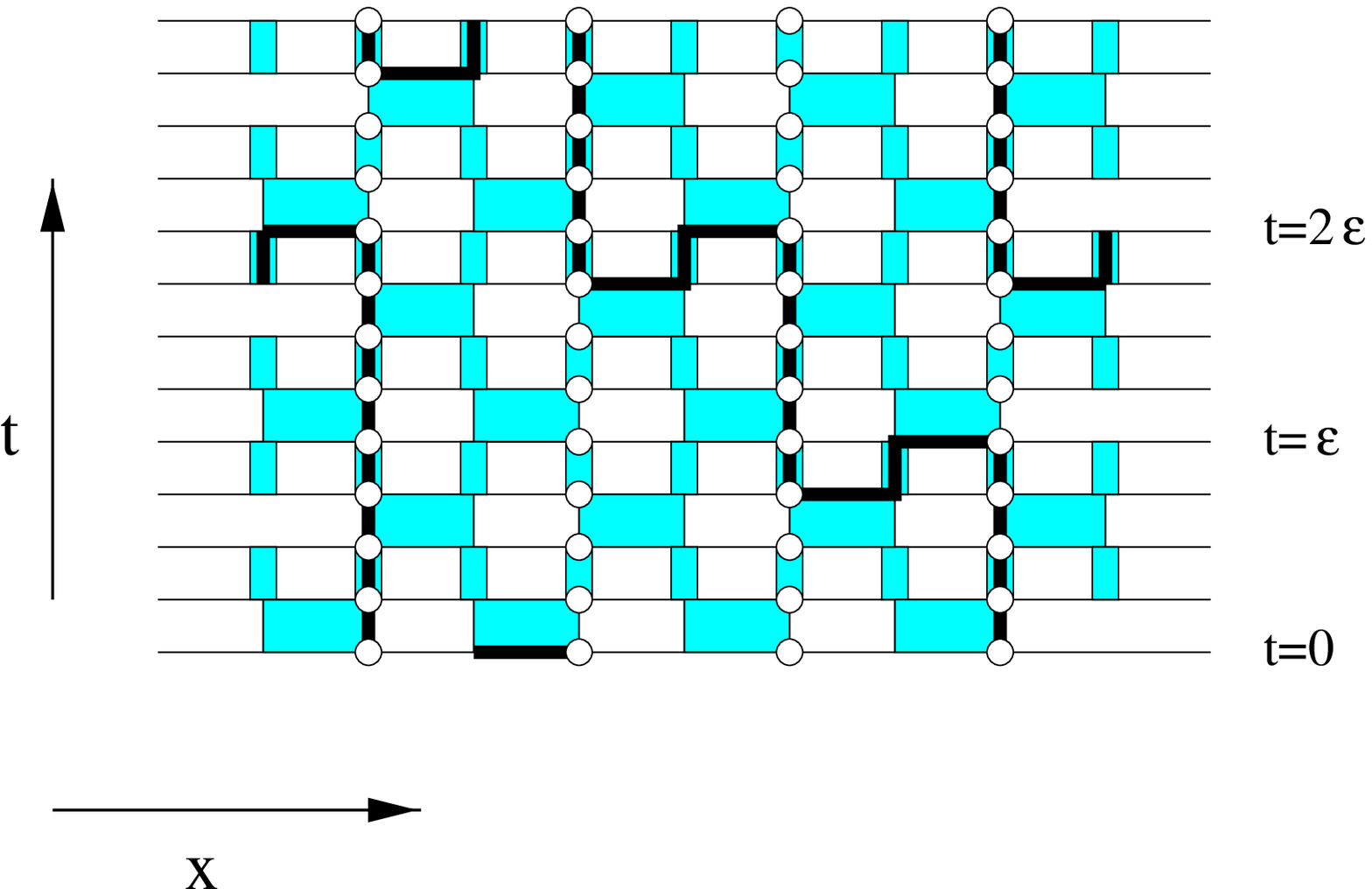,
width=3cm,angle=0,
bbllx=50,bblly=50,bburx=225,bbury=337}
}
\vspace{1.5cm}
\figcaption{\label{cconf}
The figure on the left shows a {\sl connected} fermion 
world-line configuration in one space and one time dimension. The figure
on the right is obtained after flipping a {\sl cluster} (which is shown) 
from the configuration on the left. The resulting new {\sl connected} fermion 
world-line configuration is the ``reference'' configuration of the model.}
\end{figure}

\begin{figure}[tbh]
\hbox{
\hspace{2.5cm}
\epsfig{file=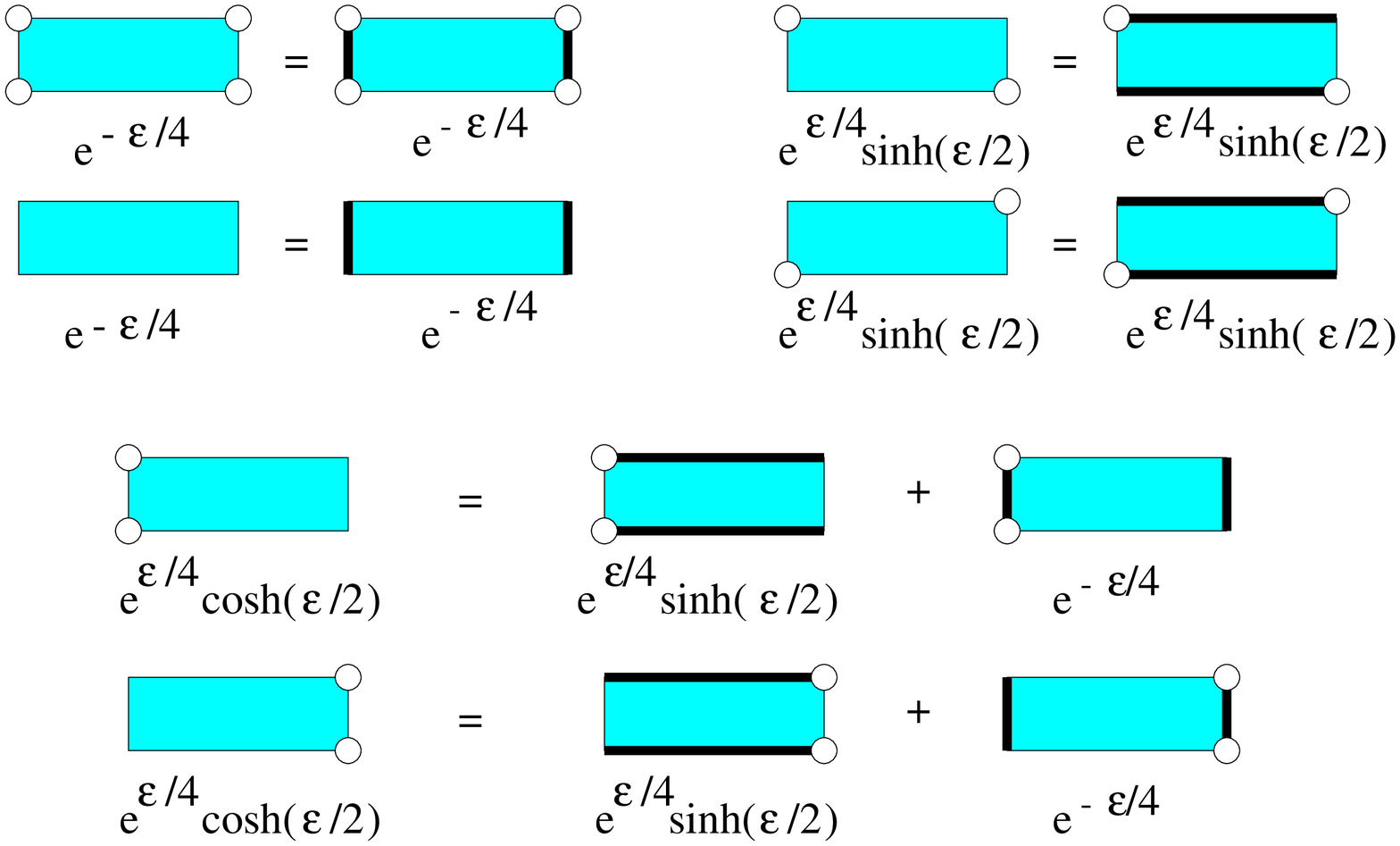,
width=3.2cm,angle=0,
bbllx=50,bblly=50,bburx=225,bbury=337}
}
\vbox{
\hbox{
\hspace{5.3cm}
\epsfig{file=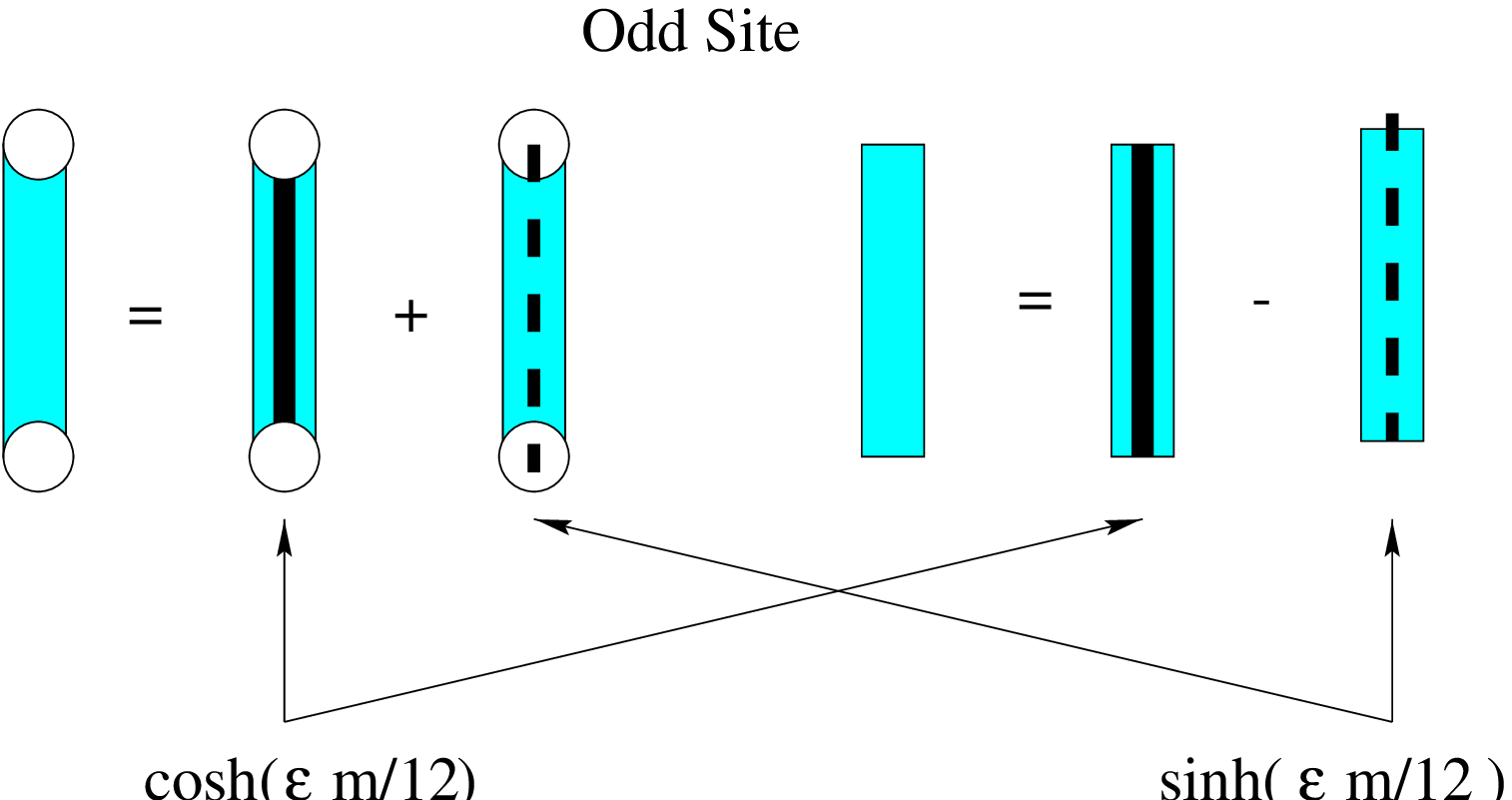,
width=3.2cm,angle=0,
bbllx=50,bblly=50,bburx=225,bbury=337}
}
\hbox{
\hspace{5.3cm}
\epsfig{file=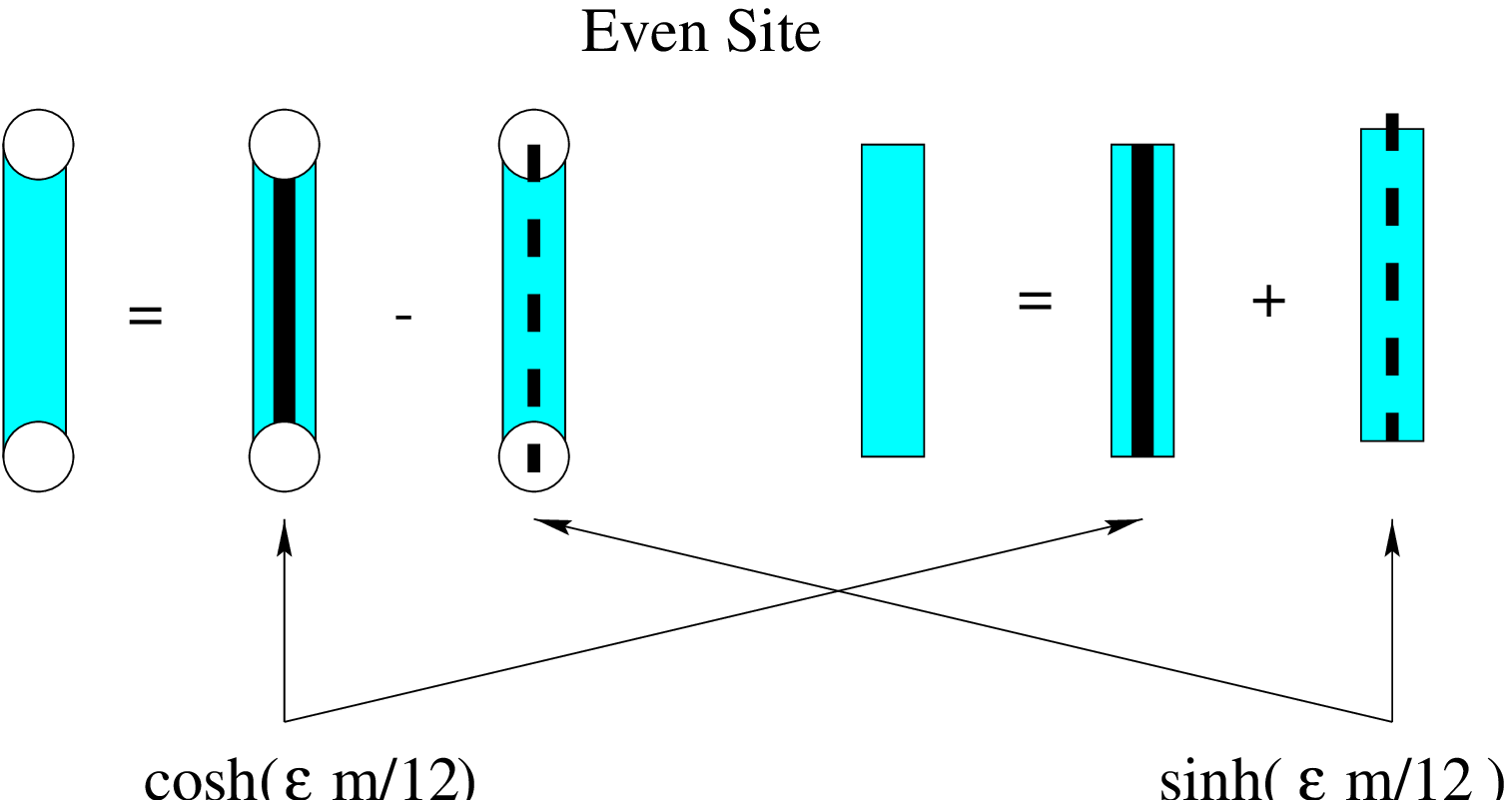,
width=3.2cm,angle=0,
bbllx=50,bblly=50,bburx=225,bbury=337}
}
}
\vspace{1.2cm}
\figcaption{\label{btmat} The figure shows how to rewrite the weight of
each configuration of figure \ref{tmat}, in terms of new configurations 
that have bonds in addition to fermions.}
\end{figure}

\section{Bond Variables and Meron Clusters}

  When the partition function is written as a sum over fermion
world-line configurations, the Boltzmann weight of each
configuration is not guaranteed to be positive definite. This leads
to the well known fermion sign problem. It was discovered recently 
that sometimes it is possible to solve this problem when 
the Boltzmann weight of each fermion world line configuration is 
written as a sum of Boltzmann weights describing new type of 
configurations, referred to as ``connected'' fermion world line 
configurations, obtained by the introduction of {\sl bond} 
variables that establish connections with in the lattice. An
example of this new type of configuration is shown in figure 3. 
In a class of models,
\clearpage
\noindent 
including the one considered in this article, it is possible to 
regroup these new type of configurations 
such that the sum of the weights of each group is positive or zero, thus 
canceling all negative weights. The regrouping is accomplished by 
identifying all possible connected fermion world-line configurations 
obtained by flipping clusters of connected lattice sites. By flipping 
a lattice site one means that if the site is filled (with a fermion) it 
is emptied and vice versa. The result of a cluster flip is also shown 
in Figure 3.

In order to calculate the Boltzmann weight of every new type of
configurations each interaction plaquette of in the old
configuration given in figure 2 is rewritten as a sum of 
new configurations that involve bonds in addition to the fermions. 
The sign factors $\Sigma$ are ignored at this step and 
will be considered during the regrouping step.
Further, the bonds are introduced so that they obey the property 
of a {\sl flip symmetry}. This means that the magnitude of the Boltzmann 
weight of connected fermions world line configurations do not change
when connected sites are flipped. This property will be useful to 
identify configurations with equal weight but opposite signs
during the regrouping step. For the two site interactions the 
weights of the new bond variables is shown at the top of figure 4.
For the single site interaction due to the mass terms one has to 
distinguish between even and the odd sites since the weights differ 
on these two sites. However, now by introducing two types of bonds 
the property of flip symmetry can be maintained as shown in the bottom 
part of figure 4. The difference between the two bonds is that when the 
sites connected by the ``dashed'' 
bond are flipped, an extra sign change has to be taken into account. 
These signs must be multiplied to the $\Sign[n]$ and the overall 
sign should be taken into account at the regrouping stage.

The above procedure gives a new expression for the path integral 
in terms of connected fermion world-line configurations.
\begin{equation}
Z_f = \sum_{{\cal C}} \Sign[{\cal C}]\; W[{\cal C}]
\end{equation}
Each configuration ${\cal C}$ now defines a set of clusters 
in addition to the fermion occupation numbers 
as shown in figure \ref{cconf}. For the present model the
clusters form closed loops. The weight $W[{\cal C}]$ of each 
configuration can be read off using the rules of figure \ref{btmat}. 
The $\Sign[{\cal C}]$ is
the product of $\Sign[n]$ for the old configuration obtained by
ignoring the bond and the extra negative sign factors that may arise 
from filled dashed-bonds on even sites and empty dashed-bonds on odd sites. 
A great simplification occurs when the flip symmetry imposed during
the bond introduction is used. If a configuration ${\cal C}$ has 
$N_{\cal C}$ clusters, then all the $2^{N_{\cal C}}$ configurations, 
obtained by flipping the clusters independently, have the
same weight $W[{\cal C}]$. This degeneracy can be used to regroup the 
configurations and write
\begin{equation}
\label{clustpart}
Z_f = \sum_{{\cal C}} \;{\bSign}\; W[{\cal C}]
\end{equation}
where $\bSign$ is the average of $\Sign[{\cal C}]$ over these 
$2^{N_{\cal C}}$ configurations. The usefulness of eq. (\ref{clustpart}) 
can be appreciated when $\Sign[{\cal C}]$ obeys the following two properties:
\begin{itemize}
\item[(1)] $\Sign[{\cal C}] = \prod_{i=1}^{N_{\cal C}} \Sign[c_i]$, where
$\Sign[c_i]$ is the sign of the $i$th cluster configuration labeled $c_i$.
\item[(2)] For every configuration the individual clusters can be 
flipped to a reference configuration $c_i^{\rm ref}$ such that 
$\Sign[c_i^{\rm ref}] = 1$.
\end{itemize}
It is then easy to show that $\bSign=0$, if there is at least a single cluster 
whose flip changes the sign of the configuration, i.e., there is a cluster 
$c_i$ such that $\Sign[c_i^{\rm non-ref}] = -1$. Such clusters are referred 
to as {\sl merons}. A configuration with no merons will then have $\bSign=1$. 
There are models in which $\Sign[{\cal C}]$ has the above two properties
and the present model falls in this class. Although the first property of 
$\Sign[{\cal C}]$ is difficult to establish, it is easy to show that every 
cluster in the present model can always be flipped such that all even sites 
are empty and odd sites are filled in the present model. This leads to the 
staggered configuration shown on the right in figure \ref{cconf}.

The new expression for the partition function derived in eq. (\ref{clustpart})
shows that fermion dynamics in the present model is equivalent to a 
statistical mechanics of clusters. Each cluster has two degrees of freedom 
related to the fermions. The Pauli-exclusion principle is encoded in the 
fact that meron clusters contribute zero weight to the partition function. 
In order to determine if a cluster is a meron or not, one has to understand 
its topology. If $n_{\rm h}$ is the number of hops of the cluster to the 
neighboring lattice site on the same time slice, $n_\eta$ is the number of 
local signs that the cluster encounters during the hops, $n_w$ is the 
temporal winding and $n_d$ is the number of dashed bonds in the cluster, 
then the cluster is a meron if $(n_w+n_{\rm h}/2+n_d+n_\eta)$ is an even 
integer. Cluster algorithms based on the concept of merons was originally 
invented in \cite{Bie95} and are referred to as the {\sl meron cluster 
algorithms}.

\section{Algorithm and Results}

\begin{figure}[ht]
\begin{center}
\hbox{
\hspace{2cm}
\epsfig{file=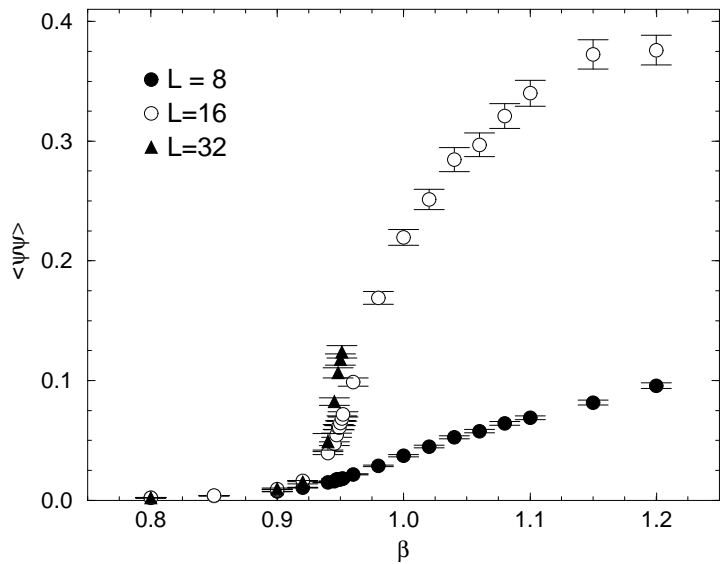,
width=3.5cm,angle=0,
bbllx=50,bblly=50,bburx=225,bbury=337}}
\figcaption{\label{results} The figure shows the chiral condensate 
$\langle \bar\psi\psi\rangle$ as a function of the inverse
temperature $\beta$ at $m =0.001$ for various spatial volumes.}
\end{center}
\end{figure}

It is easy to construct cluster algorithms that generate clusters
with weight $W[{\cal C}]$ obtained using the rules given in
figure \ref{btmat}. This makes it possible to simulate the modified
model obtained by ignoring the sign factors. The partition function of
the resulting model, denoted by $Z_b$, can be expressed as a sum over 
contributions from various meron sectors, i.e., $Z_b = \sum_N Z_N$ 
where $N$ denotes the number of merons in a cluster configuration. On 
the other hand the actual partition function $Z_f = Z_0$. Since a typical 
configuration of $Z_b$ is filled with merons, it is exponentially
difficult to find zero meron sectors and the cluster algorithm is 
still inefficient for simulating the dynamics of $Z_f$. A solution to
the problem of avoiding configurations with meron clusters is necessary.
Fortunately, using a local Metropolis accept-reject step, it is possible 
to introduce a re-weighting factor that suppresses higher meron sectors. 
For example the Metropolis step can modify the weight of cluster 
configurations to $W[{\cal C}]/p^N$, where $p>1$ and $N$ 
represents the number of merons in the configuration. This modifies 
$Z_b$ but leaves $Z_f$ unchanged! This Metropolis step then increases
the efficiency of the algorithm by an exponential factor.

A variety of quantities can be calculated with the new algorithm. 
The first step, however, is to find expressions for these quantities 
in terms of cluster variables. Operators that are diagonal in the occupation
number basis are easy to calculate, although more general operators
can also be evaluated. For some quantities it is possible to find analytic 
expressions for the average over all the cluster flips, which are referred 
to as improved estimators. The chiral condensate in the present problem,
which is also the order parameter for the chiral phase transition,
is one such quantity. In the operator language the condensate is
defined as
\begin{equation}
\langle \bar\psi\psi \rangle = 
\frac{1}{V}\frac{
\mathrm{Tr} [\mathrm{e}^{-\beta H} {\cal O}]}
{\mathrm{Tr} [\mathrm{e}^{-\beta H}]}
\end{equation}
where ${\cal O} = \sum_x p_x$. The improved estimator on the
other hand is given by
\begin{equation}
\langle \bar\psi\psi \rangle = \frac{1}{2}\frac{
\langle p\;\mathrm{Size}( c_{\mathrm{meron}})\; \delta_{N,1}\rangle }
{\langle \delta_{N,0}\rangle\;\; V\beta},
\end{equation}
where $N$ is the number of merons in the cluster configuration generated
by the algorithm and
$\mathrm{Size}( c_{\mathrm{meron}})$ is the size of
the meron cluster. The extra re-weighting factor $p$ takes into account 
the modifications due to the Metropolis step. It is easy to
show that on a finite lattice in the $m=0$ limit there are no
single meron sectors and hence the chiral condensate simply vanishes.
On the other hand, as seen in figure \ref{results}, an extremely tiny 
mass $m=0.001$ shows convincingly the effects of spontaneous symmetry 
breaking. The critical value of $\beta_c=0.948$ obtained through finite 
size scaling analysis in \cite{Sch99.2} appears consistent with the 
present results. In \cite{Sch99.2} the chiral susceptibility was measured
in the chiral limit its behavior near the transition was shown to be 
consistent with the universality class of the 3-d ising model. A more 
detailed analysis of the data from the present study will be presented 
elsewhere.

\section{Conclusions}

The fermion cluster algorithm constructed in this article is an
example of a novel way to compute fermionic path integrals. The 
sign problems that arise in the new method are solved using the
concept of a {\sl meron} which makes it easy to match all the 
configurations with negative Boltzmann weight with configurations 
with an equal but positive Boltzmann weight. Some interesting features 
emerge with the new algorithm. For the first time it is possible to 
treat fermions and bosons on an equal footing in simulations\cite{Cox99}. 
Further it is possible to approach chiral limits more easily as discussed 
here. 

A number of applications of the new techniques are presently being studied. 
For example it appears possible to find models with continuous chiral 
symmetry in which the sign problem can be completely solved. Chiral symmetry 
breaking in such a system will lead to the existence of massless Goldstone 
particles in the chiral limit. Apart from clarifying issues of universality 
the presence of the low mass particles can help in studying resonance physics 
that become difficult with conventional algorithms. Applications to 
non-relativistic many fermion problems is another topic where the present 
techniques are directly applicable. It is also likely that solutions to 
Hubbard type models will emerge with these techniques.

Adding gauge fields introduces new negative signs. It will be exciting to 
find solutions to this class of sign problems using cluster methods. There 
is evidence from strong coupling limits that a regrouping of configurations
that differ in both gauge and fermion content can help solve such problems. 
This question is presently under investigation.

\section*{Acknowledgments}

I would like to thank R. Brower, J. Cox, K. Holland, J. Osborne and 
U.-J. Wiese for collaboration and insightful discussions. The numerical 
results presented here were obtained by J. Osborne. The work was supported 
in part by a grant from the U.S. Department of Energy, Office of 
Research (DE-FG02-96ER40945).

\end{document}